\documentclass[authoryear]{elsarticle}

\title{Weak Permission is not Well-Founded, Grounded and Stable}
\author[1,2]{Guido Governatori}
\ead{g.governatori@cqu.edu.au}
\affiliation[1]{
    organization={School of Engineering and Technology, Central Queensland University},
    addressline={544 Yaamba Road},
    city={Rockhampton, QLD},
    postcode={4701},
    country={Australia}
}
\affiliation[2]{
    organization={Artificial Intelligence and Cyber Futures Institute, Charles Sturt University},
    city={Bathurst, NSW},
    postcode={4701},
    country={Australia}
}

\usepackage{amsmath}
\usepackage[helvratio=.90]{newtx}
\usepackage{microtype}
\usepackage{enumitem}
\setlist{noitemsep}

\newcommand{\obl}{\mathop{\mathsf{obl}}}
\newcommand{\perm}{\mathop{\mathsf{perm}}\nolimits}
\newcommand{\naf}{\mathop{\mathsf{not}\;}}
\newcommand{\dop}{\mathop{\mathsf{dop}}}
\newcommand{\non}{\mathnormal{\sim}}
\newcommand{\sub}{\mathop{\mathrm{Sub}}}

\newtheorem{theorem}{Theorem}
\newdefinition{definition}{Definition}
\newdefinition{example}{Example}
\newproof{proof}{Proof}

\begin{document}

\begin{abstract}
	We consider the notion of weak permission as the failure to conclude that the opposite obligation. We investigate the issue from the point of non-monotonic reasoning, specifically logic programming and structured argumentation, and we show that it is not possible to capture weak permission in the presence of deontic conflicts under the well-founded, grounded and (sceptical) stable semantics.
\end{abstract}


\maketitle

\section{Introduction}
\label{sec:intro}

Deontic Logic is the branch of logic that investigates the logical behaviour of obligations, permissions, prohibitions and related notions. Most Deontic Logics take obligation as primitive and leave the others as derived from obligations. On the other hand, normative reasoning/legal theory identifies two different notions of permission: \emph{Strong Permission} and \emph{Weak Permission}.  While the definitions of the types of permission vary, and other notions of permission have been proposed (for a discussion on the topic, see \cite{handbook:permission}), often strong permission is taken as a derogation to a prohibition or the obligation to the contrary, and we have a weak permission when we fail to obtain the obligation of the contrary.  

Another way to look at the issue is whether there are norms that explicitly permit something. If there are and the norms are effective, then we obtain an explicit (strong) permission. 

Thus, we have a weak permission if no norms make the opposite obligatory. But what about if there are norms for the opposite that are not effective? Here we have two cases. The first case is when there is a norm for the opposite, but the norm is not applicable (namely, the conditions to apply the norm do not hold). Effectively, we can treat this situation as one without norms. The second case is when there are norms for the obligation of the opposite, but these norms conflict with some other norms for the conclusion.  So, the basic situation here is that we have a norm for $\obl(a)$ ($a$ is obligatory) and a second norm for $\obl(\neg a)$ ($a$ is forbidden, or $\neg a$ is obligatory), and there is no \emph{prima facie} mechanism to solve the conflict.  There are a few options. The first is to recognise that there is a conflict and prevent any conclusion following from the conflict. In this case, we cannot assert that we have a weak permission. The second option is to endorse the conflict as the failure to obtain the obligations and to accept the corresponding weak permissions. For example, \citet{icail23} argue that for some legal applications, we stop the conflict, but we use it (in this case we use the failure to derive an obligation) to obtain further conclusions.  

Let us illustrate the situation with an example inspired by a real-life case.

\begin{example}
In 2019, the Sea Watch 3, an NGO vessel, was on a mission to rescue migrants in the Mediterranean Sea.  However, the Italian Government issued a decree that banned the Sea Watch 3 from rescue operations (for migrants) in the Italian Contiguous Zone waters. 
\begin{gather*}
    r_1\colon\mathit{distress}, \mathit{proximity} \Rightarrow \obl(\mathit{assistance})\\
    r_2\colon\mathit{SeaWatch}, \mathit{migrants}, \mathit{ItalianContiguousZone} \Rightarrow \obl(\neg\mathit{assistance})
\end{gather*}  
Maritime Law stipulates that a vessel in proximity to a distress vessel must 
assist the vessel in distress (rule $r_1$).
On the other hand, $r_2$ encodes the prohibition on rescuing migrants in Italian Contiguous Zone waters. Moreover, Maritime Law requires vessels permitted to refrain from assisting vessels in distress to alert the closest relevant authorities and keep clear of the rescue area.
\[
 r_3\colon \perm(\neg\mathit{assistance}) \Rightarrow \obl(\mathit{alertAuthorities}) \& \obl(\mathit{keepClear})
\]
Suppose a migrant vessel is in distress in the Italian Contiguous Zone, and Sea Watch 3 is nearby. Does Sea Watch 3 have the obligation to alert the authorities and keep clear of the rescue area?  
Here, we have a conflict over two opposite norms, and it is unclear which one takes precedence over the other.\footnote{In response to the Sea Watch 3 incident, an Italian Tribunal established that International Maritime Law prevailed over the Italian Government Decree for the specific case. However, International Maritime Law scholars debated over the proper course of action.} Also, suppose that Sea Watch 3 alerted the competent authorities but remained in the operation area (and eventually assisted with the rescue operation). Do the Sea Watch 3 actions contravene the norms?   As we said, there are two options.  
The permission does not follow from the conflict. In this case, the Sea Watch is not required to leave the area, and it does not contravene the norms. In the second approach, the permission from the conflict follows (as a weak permission). Leaving the area is a legal requirement, and Sea Watch 3 does not comply with the norms.

What about if weak permission does not follow from a conflict? We can consider an alternative formulation of the last provision. Assume that it stipulates that vessels for which the obligation to assist does not hold must keep clear of the rescue area. 
\[
  r'_3\colon \naf\obl(\mathit{assistance}) \Rightarrow \obl(\mathit{alerAuthorities})\&\obl(\mathit{keepClear})
\]
Given the conflict, we can argue that the obligation to offer assistance does not hold, and the Sea Watch 3 has to follow the obligations given by $r'_3$.
\end{example}
The example above illustrates a feature of legal and normative reasoning: obligations, permissions, and prohibitions (and the lack of them) can trigger other obligations, permissions and prohibitions. Logics and formalisms for deontic reasoning must handle these cases. Furthermore, it has been argued (for example by \cite{Jones:1992}) that, in general,  deontic reasoning is an essential component of normative reasoning. At the same time, normative reasoning is by its own nature non-monotonic (specifically, defeasibility is a key part of normative and legal reasoning). Indeed, many works propose to integrate and found deontic and normative reasoning with non-monotonic reasoning. Too many works to mention, we limit ourselves to the seminal proposal by \citet{BNA} to use logic programming for modelling norms (albeit with the severe limitation of lacking deontic reasoning), the approach by \citet{Horty1993} to look at deontic logic from the non-monotonic reasoning lens, and the survey on formal argumentation approaches to legal reasoning by \cite{PrakkenSartor2015}. However, two common themes have emerged: either they incorporate deontic features or address the suitability of the underlying formalisms, logics and semantics for deontic reasoning. 

In the rest of this paper, we look at the issue of weak permission from two popular paradigms for non-monotonic reasoning and some of the most adopted semantics. More specifically, in Section~\ref{sec:lp}, we examine logic programming as formalism and well-founded \citep{vangelder1991} and stable semantics (\cite{gel88}).  Then, in Section~\ref{sec:arg}, we study how to integrate deontic reasoning in (structured) argumentation \citep{Prakken2010}, and we look at it through the lens of grounded and stable semantics for argumentation \citep{Dung1995}.  Given the close relationships between logic programming and argumentation and their semantics, we get the same result: weak permission is not supported by well-founded, grounded and (sceptical) stable semantics when there are deontic conflicts. 

\section{Logic Programming and Weak Permission}
\label{sec:lp}

The language of a deontic program is built from a set of literals, where a literal is either an atom ($l$) or its negation ($\neg l$). In addition, we extend the language with \emph{deontic literals}. The set of deontic literals is defined as
\[
\{\dop(l),\naf\dop(l)| l\in \mathrm{Lit} \}
\]
where $\dop$ is a deontic operator, more precisely $\dop\in\{\obl,\perm,\perm_w\}$. 
The language also admits negation as failure ($\naf$). Literals and deontic literal can appear in the scope of $\naf$, but negation as failure cannot appear in the scope of a deontic literal. Given a literal $l$, we use $\non l$ to denote the complement of $l$; more precisely, if $l$ is an atomic proposition, then $\non l=\neg l$. If $l$ is a negated atomic proposition $l=\neg m$, then $\non l=m$.

\begin{definition}[Program]
A \emph{program} is a set of clauses or rules, where a rule $r$ is an expression 
\[
  c \leftarrow a_1,\dots,a_m, \naf b_1,\dots,\naf b_m.
\] 
where $a_1,\dots,a_n,b_1,\dots,b_m,c$ are either literals or deontic literals, with the restriction that $c$ is not a weak permission. 
\end{definition}
\noindent
Given a rule $r$, $c$ is called the head of the rule, noted $h(r)$, and $\{a_1,\dots,a_n,b_1,\break\dots,b_m,\}$ is the body of $r$ (noted $b(r)$). We split the body into the positive part $b^+(r)=\{a_1,\dots,a_n\}$ and the negative part $b^-(r)=\{\naf b_1,\dots,\naf b_m\}$.  The head and the body can be empty. We will refer to a rule with an empty head as an \emph{integrity constraint}. 
The \emph{Herbrand Base} of a program $P$ is the set of all literals and deontic literals appearing in $P$. 

To capture deontic reasoning, we introduce deontic programs, namely, programs extended with a set of additional clauses encoding key properties of deontic reasoning. 
\begin{definition}[Deontic Program] 
A \emph{deontic program} is a program including the following clauses and integrity constraints:
\begin{gather}
	\perm(X) \leftarrow \obl(X). \label{eq:OtoP}\\
	\perm_w(X) \leftarrow \naf\obl(\non X). \label{eq:nafP}\\
	\leftarrow \obl(X), \obl(\neg X). \label{eq:OOnotIC}\\
	\leftarrow \obl(X), \perm(\non X) \label{eq:OPIC}.
\end{gather}
\end{definition}
\noindent
The rule in \eqref{eq:OtoP} corresponds to the D axioms of Standard Deontic Logic, where an obligation implies the corresponding permission. Clause \eqref{eq:nafP} establishes that we have a weak permission if we fail to derive the obligation to the contrary. The two integrity constraints ensure the (deontic) consistency of a deontic program. So, according to the first integrity constraint, no proposition can be at the same obligatory and forbidden. Similarly, according to \eqref{eq:OPIC} no literal can be obligatory (forbidden) when its opposite is permitted. 

From the first integrity constraint, we have that no program with the following two rules 
\[
 \obl(p) \leftarrow.\qquad\qquad
 \obl(\neg p)\leftarrow.
\]
is satisfiable. Accordingly, to have a satisfiable deontic conflict we have to encode the rules for obligation as follows:
\[
	\obl(p) \leftarrow \naf\obl(\neg p).\qquad\qquad
	\obl(\neg p) \leftarrow \naf\obl(p).
\]

We are now ready to introduce the semantic framework for deontic programs. We follow the presentation by \citet{caminadas2015}. While the approach is not standard, its advantage is that it provides a unified view of well-founded and stable semantics, where the well-founded semantics can be seen as a 3-valued variant of the 2-valued stable semantics as proposed by \citet{przymusinski1990}.
\begin{definition}
    A 3-valued Herbrand Interpretation $I$ of a logic program $P$ is a pair $\langle T, F\rangle$ with $T, F \subseteq HB_P$ and $T \cap F=\emptyset$. The atoms in $T$ are said to be \emph{true}, the atoms in $F$ are said to be \emph{false} and the atoms in $HB_P \backslash(T \cup F)$ are said to be \emph{undefined}.
\end{definition}

\begin{definition}\label{def:reduct}
    Let $I$ be a 3-valued Herbrand Interpretation of the logic program $P$. The \emph{reduct} of $P$ with respect to $I$ (written as $P/I$) is the logic program constructed using the following steps.
    \begin{enumerate}
        \item Starting from $P$, remove each rule $r$ from $P$ that has $\naf b_i \in b^{-}(r)$ for some $b_i \in T$;
        \item From the result of step 1, for each rule, for every $b_i \in F$, remove  $\naf b_i$ from the body of the rule;
        \item From the result of step 2, replace any remaining occurrences of  $\naf b_i$ by $\mathbf{u}$.
    \end{enumerate}
    where $\mathbf{u}$ is an atom not in $H B_P$ which is undefined in all interpretations of $P$ (a constant).
    $\Psi_P(I)=\left\langle T_{\Psi}, F_{\Psi}\right\rangle$ with minimal $T_{\Psi}$ and maximal $F_{\Psi}$ (w.r.t. set inclusion) is the unique least 3-values model of $P$ such that, for every $a \in HB_P$:
    \begin{itemize}
        \item $a \in T_{\Psi}$ if there is a rule $r' \in P/I$ with $h(r')=a$ and $b^{+}(r') \subseteq T_{\Psi}$;
        \item $a \in F_{\Psi}$ if every rule $r'\in P/I$ with $h(r')=a$ has $b^{+}(r') \cap F_{\Psi} \neq \emptyset$.
    \end{itemize}
\end{definition}

\begin{definition}[Logic Programming Semantics] 
    Let $I=\langle T, F\rangle$ be a 3-valued Herbrand Interpretation of logic program $P$.
\begin{itemize}
    \item $I$ is a partial stable (or P-stable) model of $P$ iff $\Psi_P(I)=I$.
    \item $T$ is a well-founded model of $P$ iff $I$ is a P-stable model of $P$ where $T$ is minimal (w.r.t. set inclusion) among all P-stable models of $P$.
    \item $T$ is a (2-valued) stable model of $P$ iff $I$ is a $P$-stable model of $P$ where $T \cup F=HB_P$.
    \item Let $\mathcal{I}=\{I_i=\langle T_i,F_i\rangle\colon\text{such that } T_i \text{ is a stable  model of } P\}$. $T=\bigcap_{I_i\in\mathcal{I}}T_i$ is the sceptical stable model of $P$.    
\end{itemize}
\end{definition}

\begin{definition}[Conflicted Deontic Program]
    A deontic program $P$ is \emph{conflicted} if, for a literal $l$, it contains the rules 
    \[\obl(l)\leftarrow \naf\obl(\neg l).
     \qquad\qquad
     \obl(\neg l)\leftarrow\naf\obl(l).
    \] 
    We say that $l$ is a \emph{conflicted literal}.
\end{definition}

\begin{theorem}
    Let $P$ be a conflicted deontic program, and let $l$ be a conflicted literal. Then $\perm_w(l)$ and $\perm_w(\neg l)$ are not conclusions of the program under well-founded and stable semantics.     
\end{theorem}
\begin{proof}
To prove the property we can assume, without any loss of generality, that the language is restricted to the conflicted literal $l$, the Herbrand Base consists of the deontic literals that can be built from $l$, i.e., $HB_P=\{\obl(l),\obl(\neg l), \perm(l), \perm(\neg l),\perm_w(l), \linebreak \perm_w(\neg l)\}$, and the program consists of the two rules such that $l$ is conflicted plus the instantiation of clauses \eqref{eq:OtoP}--\eqref{eq:OPIC}. 

First, we show that $I=\langle\emptyset,\emptyset\rangle$, is a partial stable model.  By Definition~\ref{def:reduct} $P/I$ contains the following rules (for the sake of clarity and conciseness we removed the instances of the integrity constraints):
\begin{align*}
    \perm(l)&\leftarrow \obl(l). &  \perm_w(l) &\leftarrow \mathbf{u}. & \obl(l)&\leftarrow \mathbf{u}.\\
    \perm(\neg l)&\leftarrow \obl(\neg l). & \perm_w(\neg l) &\leftarrow \mathbf{u}.  & \obl(\neg l)& \leftarrow \mathbf{u}.
\end{align*}    
It is easy to verify that $T_\Psi=\emptyset$ and $F_\Psi=\emptyset$. Clearly, $T_\Psi=\emptyset$ is minimal. Let us show that $F_\Psi=\emptyset$ is maximal. Given the integrity constraints \eqref{eq:OOnotIC} and \eqref{eq:OPIC} it is not possible that $\obl(l)$ and $\obl(\neg l)$ are true in the same interpretation; similarly for $\obl(l)$ and $\perm(\non l)$, and by rule \eqref{eq:nafP} $\obl(l)$ and $\perm_w(\neg l)$ cannot be both true. 

Suppose that there is an interpretation $I'=\langle \emptyset,F'\neq\emptyset\rangle$. Moreover, assume $\obl(l)\in F$. Then $P/I'$ contains the rule $\obl(\neg l)\leftarrow.$ Thus, $\obl(\neg l)\in T_\Psi$, and so,$I'\neq\Psi_P(I')$. Suppose $\perm(l)\in F'$, given the rule $\perm(l)\leftarrow\obl(l).$ $\obl(l)$ must be in $F'$ as well, and we can repeat the argument above. Finally, if $\perm_w(l)$ were in $F'$, then $P/I'$ should either not contain any rule for $\perm_w(l)$, but this means that $\obl(\neg l)\in T$ (which is not the case); or $\perm_w(l)\leftarrow\mathbf{u}\in P/I'$, and then $\mathbf{u}\in F'$; however, as stipulated, $\mathbf{u}$ does not belong to $HB_P$, and so, it is not in $F'$.  The proof for deontic literals based on $\neg l$ is the same.  Accordingly, we have shown that $F_\Psi=\emptyset$ is maximal, and so, $T_\Psi$ is the well-founded model of $P$, and $\perm_w(l),\perm_w(\neg l)\notin T_\Psi$. 

For the stable semantics, we are looking for interpretations where the elements of the (deontic) Herbrand Base are distributed over the set of true and false literals. As noted above, if one interpretation, $\obl(l)$ and $\obl(\neg l)$ cannot be true in the same interpretation. So, one of them must be false, let us say $\obl(l)\in F$, then $P/I$ contains the rule $\obl(\neg l)\leftarrow.$ Thus, the extension $\obl(\neg l)\in T$. Accordingly, we have two interpretation $I_1=\langle T_1,F_1\rangle$, and $I_2=\langle T_2,F_2\rangle$ where $\obl(l)\in T_1$ and $\obl(\neg l)\in T_2$.  Hence, the two reducts are 
\begin{align*}
    P/I_1=\{\perm(l)&\leftarrow\obl(l). & 
            \perm(\neg l)&\leftarrow\obl(\neg l). &
            \perm_w(l)&\leftarrow. &
            \obl(l)&\leftarrow.\}\\
    P/I_2=\{\perm(l)&\leftarrow\obl(l). & 
            \perm(\neg l)&\leftarrow\obl(\neg l). &
            \perm_w(\neg l)&\leftarrow. &
            \obl(\neg l)&\leftarrow.\}
\end{align*}
and the complete interpretations are 
\begin{gather*}
    T_1=F_2=\{\obl(l), \perm(l),\perm_w(l)\}\\
    F_1=T_2=\{\obl(\neg l), \perm(\neg l),\perm_w(\neg l)\}
\end{gather*}
Thus, we have two stable models, one containing $\perm_w(l)$ and the other containing $\perm_w(\neg l)$. 
Therefore, $\perm_w(l)$ and $\perm_w(\neg l)$ are not sceptical conclusions under the stable semantics.
\end{proof}

\section{Argumentation and Weak Permission}
\label{sec:arg}

For the language for our deontic argumentation framework we use the same language $\mathcal{L}$ as the previous part, but we do not admit negation as failure.  Arguments are built from rules where a rule has the following format:
\[
a_1,\dots a_n \Rightarrow c
\]
where $\{a_1,\dots,a_n\}$ is a (possibly empty) set of literals and deontic literal, and $c$ is either a literal or deontic literal but not a weak permission (i.e., $c\neq\perm_w(l)$, otherwise it would be an explicit permission).

\begin{definition}
A \emph{Deontic Argumentation Theory} is a structure 
\[(F,R)\]
where $F$ is a (finite and possibly empty) set of literals (the fact or assumption of the theory), and $R$ is a (finite) set of rules. 
\end{definition}
The key concept of an argumentation theory is the notion of an argument. Definition~\ref{def:argument} below defines what an argument is. Each argument $A$ has associated to it, its conclusion $C(A)$ and its set of sub-arguments $\sub(A)$.
\begin{definition}
    \label{def:argument}
Given a Deontic Argumentation Theory $(F,R)$,  $A$ is an \emph{argument} if $A$ has one of the following forms: 
\begin{enumerate}
    \item $A=\perm_w(l)$ for any literal $l\in\mathcal{L}$, the conclusion of the argument $C(A)=\perm_w(l)$, and $\sub(A)=\{A\}$.
    \item $A=a$ for $a\in F$; $C(A)=a$ and $\sub(A)=\{A\}$. 
    \item $A=A_1,\dots,A_n\Rightarrow c$, if there is a rule $a_1,\dots,a_n\Rightarrow c$ in $R$ such that for all $a_i\in\{a_1,\dots, a_n\}$
    there is an argument $A_i$ such that $C(A_i)=a_i$; $C(A)=c$ and $\sub(A)= \{A\}\cup\sub(A_1)\cup\dots\cup\sub(A_n)$.
    \item $A=B \Rightarrow \perm(l)$, if $B$ is an argument such that $C(B)=\obl(l)$; $C(A)=\perm(l)$, and $\sub(A)=\{A\}\cup\sub(B)$.
\end{enumerate}    
\end{definition}
\noindent
Condition 1) encodes the idea that weak permission is the failure to derive an obligation to the contrary (more on this when we discuss the notion of attack between arguments). Thus, by default, every literal is potentially weakly permitted, and we form an argument for this type of conclusion. 
Condition 2) gives the simplest form of an argument. We have an argument for $a$ if $a$ is one of the assumptions/facts of a case/theory. 
Condition 3) allows us to form arguments by forward chaining rules. Thus, we can form an argument from a rule, if we have arguments for all the elements of the body of the rule. The way the condition is written allows us to create arguments from rules with an empty body.  Finally, condition 4) corresponds to the D axiom of Standard Deontic Logic.

\begin{definition}[Attack]
  Let $A$ and $B$ be arguments. $A$ \emph{attacks} $B$ ($A > B$) iff
  \begin{enumerate}
    \item $B=\perm_w(l)$ and $C(A)=\obl(\non l)$;
    \item $\exists B'\in\sub(B)$, $C(B)=l$ and $C(A)=\non l$;
    \item $\exists B'\in\sub(B)$, $C(B')\in\{\obl(l),\perm(l)\}$ and $C(A)=\obl(\non l)$;
    \item $\exists B'\in\sub(B)$, $C(B')=\obl(l)$, and $C(A)=\perm(\non l)$. 
  \end{enumerate}
  We use $\mathcal{A}$ to denote the set of all arguments of the theory. 
\end{definition}
As we alluded to above and as we have seen in Section~\ref{sec:lp}, the idea of weak permission is the negation as failure of the obligation to the contrary. Thus, in Definition~\ref{def:argument}, we create an argument for the weak permission for any literal $l$; however, this argument is attacked by any argument for $\obl(\non l)$ (condition 1. above and notice this is the only case where the attack is not symmetrical). The rest of the conditions define an attack when the two arguments have opposite conclusions: condition 2 covers the case of plain literals, while conditions 3 and 4 are reserved for deontic literals. Specifically, we have opposite deontic conclusions when one of the two is an obligation for a literal, and the other is either an obligation or a permission for the opposite literal. Furthermore, an argument attacks another argument when the conflict is on the conclusion of the second argument (this corresponds to the notion of rebuttal) or when there is a conflict with one of the sub-arguments of the attacked argument (known as undercutting attack).

\begin{definition}[Dung Semantics] 
    Let $(F,R)$ be a Deontic Argumentation Theory, and $S$ be a set of arguments. Then:
\begin{itemize}
\item $S$ is conflict free iff $\forall X, Y \in S: X \not> Y$.
\item $X \in \mathcal{A}$ is acceptable with respect to $S$ iff $\forall Y \in \mathcal{A}$ such that $Y>X$ $: \exists Z \in S$ such that $Z>Y$.
\item $S$ is an admissible set iff $S$ is conflict free and $X \in S$ implies $X$ is acceptable w.r.t. $S$.
\item $S$ is a complete extension iff $S$ is admissible and if $X \in \mathcal{A}$ is acceptable w.r.t. $S$ then $X \in S$.
\item $S$ is the grounded extension iff $S$ is the set inclusion minimal complete extension.
\item $S$ is a stable extension iff $S$ is conflict free and $\forall Y\notin S$, $\exists X\in S$ such that $X>Y$.
\end{itemize}
\end{definition}

\begin{definition}[Justified Argument]
    Let $D=(F,R)$ be a Deontic Argumentation Theory, an argument $A$ is \emph{sceptically justified} under a semantic $T$ iff $A\in S$ for all sets of arguments $S$ that are an extension under $T$.
\end{definition}

\begin{definition}[Justified Conclusion]
    A literal or a deontic literal $l\in\mathcal{L}$ is a \emph{Justified conclusion} under a semantics $T$ iff for every extension $S$ under $T$, there is an argument $A$ such that $A\in S$ and $C(A)=l$
\end{definition}

Let us consider a Deontic Argumentation Theory where $F=\emptyset$ and $R$  contains the two rules 
\[
r_1: {} \Rightarrow \obl(a) \qquad \qquad r_2: {} \Rightarrow \obl(\neg a)
\]
This theory has the following arguments:
\begin{align*}
A_1&: \perm_w(a) &
A_3&: {}\Rightarrow \obl(a) &
A_5&: A_3 \Rightarrow \perm(a)\\
A_2&: \perm_w(\neg a) &
A_4&: {}\Rightarrow \obl(\neg a) &
A_6&: A_4 \Rightarrow \perm(\neg a)
\end{align*}
For the attack relation, we have the following instances
\begin{align*}
    A_3 > A_2 && 
    A_3 > A_4 &&
    A_3 > A_6 &&
    A_5 > A_4\\
    A_4 > A_1 &&
    A_4 > A_3 &&
    A_4 > A_5 &&
    A_6 > A_3
\end{align*}
It is easy to verify that $\{\}$ is a complete extension (and trivially, it is the minimal complete extension w.r.t. set inclusion). Thus, it is the grounded extension of the theory. Accordingly, there is no argument in the grounded extension such that its conclusion is either $\perm_w(a)$ or $\perm_w(\neg a)$. Hence, $\perm_w(a)$ and $\perm_w(\neg a)$ are not justified conclusions under the grounded semantics. 

When we consider the stable semantics, we have the following two extensions:
\[
    \{A_1, A_3, A_5\} \qquad \qquad
    \{A_2, A_3, A_6\}
\]
Clearly, $\perm_w(a)$ is a conclusion of the first extension but not of the second one; conversely, $\perm_w(\neg a)$ is a conclusion of the second extension but not of the first one. Consequently, $\perm_w(a)$ and $\perm_w(\neg a)$ are not justified conclusions under the stable semantics.

The above example should suffice to show that weak permission is not supported by grounded and stable semantics when a deontic conflict exists: we have a scenario where we fail to conclude that an obligation, but at the same time, we cannot conclude the weak permission of the opposite.  However, we can generalise the result; the result holds for any theory with a conflict between two applicable obligation rules. This is formalised by the following definition. 

\begin{definition}
    A Deontic Argumentation Theory is \emph{conflictual} when it contains a pair of rules $b_1,\dots,b_n\Rightarrow\obl(c)$ and $d_1,\dots d_m \Rightarrow \obl(\neg c)$,
    such that there are arguments $B_i$ with conclusion $b_i$, $1\leq i\leq n$, and $D_j$ with conclusion $d_j$, $1\leq j\leq m$. We will call $c$ the \emph{conflicted} literal.
\end{definition}

\begin{theorem}
   Let $D=(F,R)$ be a conflictual theory. For any conflicted literal $l$, $\perm_w(l)$, $\perm_w(\neg l)$ are not justified conclusions under grounded and stable semantics.    
\end{theorem}
\begin{proof}
Let $l$ be a conflicted literal. By the definition of argument, the theory $D$ and the fact that the theory is conflicted, we have the following arguments: 
\begin{align*}
    A_1&: \perm_w(l) &    A_3&: B_1,\dots, B_n \Rightarrow\obl(l)\\
    A_2&: \perm_w(\neg l) &
    A_4&: D_1,\dots, D_m \Rightarrow\obl(\neg l)
\end{align*}   
such that $A_3 > A_2$ and $A_4 > A_1$. Given that there are arguments attacking $A_1$ and $A_2$, these two arguments are not in the minimal complete extension. Accordingly, $\perm_w(l)$ and $\perm(\neg l)$ are not justified conclusions under the grounded semantics.  

Given the attack relationship among $A_1$, $A_2$, $A_3$ and $A_4$, we can conclude that there are at least two extensions $E_1$ and $E_2$ such that $A_1,A_3\in E_1$ and $A_2,A_4\notin E_1$, and $A_2,A_4\in E_2$ and $A_1,A_3\notin E_2$. Hence, there is an extension where no argument has $\perm_w(l)$ as its conclusion, and there is an extension where no argument has $\perm_w(\neg l)$ as its conclusion. Therefore, $\perm_w(l)$ and $\perm_w(\neg l)$ are not justified conclusions under the stable semantics. 
\end{proof}

\section{Summary and Conclusions}
\label{sec:conclusions}
We investigated the issue of weak permission, defined as the lack of the obligation to the contrary, in the context of some forms of non-monotonic reasoning (specifically, logic programming and structured argumentation) and some of the most adopted semantics (well-founded, grounded and stable). We proved that when the failure to obtain an obligation depends on an unsolved deontic conflict, the corresponding weak permission is not a conclusion under the well-founded, grounded and (sceptical) semantics. Accordingly, approaches to deontic reasoning adopting such semantics cannot offer a proper model of full deontic reasoning. 

\bibliographystyle{elsarticle-harv}
\bibliography{weaknotwell}

\end{document}